\pdfoutput=1  
\documentclass[%
 reprint,
 amsmath,amssymb,
 aps,
]{revtex4-2}

\usepackage{graphicx}
\usepackage{dcolumn}
\usepackage{bm}
\usepackage{hyperref}
\usepackage[mathlines]{lineno}

\begin{document}

\preprint{APS/123-QED}

\title{Non-monotonic hydration dependence of ionic transport\\ in atomically-thin MnO$_2$ sheets }

\author{Ata Utku Özkan}
\affiliation{Institute of Materials Science and Nanotechnology, Bilkent University, Ankara 06800, Turkiye}
\affiliation{Bilkent National Nanotechnology Research Center - UNAM, Ankara 06800, Turkiye}
\author{T. Serkan Kas\i rga}
\email{Contact author: kasirga@unam.bilkent.edu.tr}
\affiliation{Department of Physics, Middle East Technical University, Ankara 06800, Turkiye}
\affiliation{Bilkent National Nanotechnology Research Center - UNAM, Ankara 06800, Turkiye}
\author{Aykut Erba\c{s}}
\email{Contact author: aykut.erbas@unam.bilkent.edu.tr}
\affiliation{Institute of Materials Science and Nanotechnology, Bilkent University, Ankara 06800, Turkiye}
\affiliation{Bilkent National Nanotechnology Research Center - UNAM, Ankara 06800, Turkiye}


\date{\today}

\begin{abstract}
 The interlamellar space in van der Waals materials has served as a test bed for studies of strongly confined fluid and ion transport. Despite theoretical and experimental progress, it remains unclear how electrostatic correlations arising from interactions between the confined, hydrated ions and the substrate layer can be reconciled with non-equilibrium ionic transport and the substrate's morphological response. Here, we investigate field-driven ionic conduction in sodium- and water-intercalated layered manganese oxides (MnO$_2$), a self-intercalated metal oxide nanofluidic system, as a model intrinsically-intercalated van der Waals solid, using nonequilibrium all-atom molecular dynamics simulations that explicitly capture ion–water correlations and layer morphology. 
We demonstrate that electric-field bias induces spontaneous nanoscale segregation of water within the interlayer space, producing coexisting hydrated and weakly hydrated ionic domains coupled to local lattice distortions. This feedback between hydration and morphology gives rise to heterogeneous transport pathways and leads to a non-monotonic dependence of ionic conductivity on water content. Suppressed conduction arises either from layer collapse at low water levels or from ionic hop blockage by excess water at high water levels. Concurrently, ionic transport exhibits a maximum at intermediate levels of intercalated water, where less hydrated ions can move at the boundaries of strongly hydrated ion clusters.
 While such non-monotonicity could explain the experimentally observed memristive response of MnO$_2$ single crystals, these findings also provide a molecular-level mechanism linking intercalated water levels to ionic metal oxide nanofluidic systems, suggesting general design principles for robust, water-assisted ionic conductors.
 \end{abstract}

\maketitle

\section{Introduction}

The transport of ionic charge in nanoscale confinement exhibits a broad range of new phenomena that are otherwise not observed in bulk materials, leading to a wide range of emerging technologies, including electrochemical energy storage, neuromorphic devices, catalysis, and ionic information processing~\cite{watanabe2017ionic,JOSHI2025,Kulathuvayal2023,zhou2021,Robin2021}. When ionic motion occurs within angstrom-scale channels under non-equilibrium conditions, classical continuum descriptions of electrolyte transport frequently break down due to strong electrostatic correlations, altered solvation structures, and dynamic coupling between mobile ions and confining interfaces~\cite{Netz2020Review,Lyderic2023Review,Chen2023NanoLetters}.
Importantly, a largely overlooked yet fundamental degree of freedom is the response of the confining surface to the applied external field, particularly in systems where a nanoconfined fluid, often water, and ionic species actively stabilize and modulate the morphology of the confining substrate~\cite{nature_shelby2021}.

Among such ionic systems, in layered metal oxides (e.g., MnO$_2$) the electrostatic charge balance between charged host layers  and oppositely charged interlayer ions (e.g., potassium or sodium) stabilizes the lattice while simultaneously permitting ionic motion under sub-nanometer confinement~\cite{serkan_hoca2021,Gogotsi2017,Shan2019}. Importantly, ion-intercalated metal-oxides such MnO$_2$, particularly birnessite-type structures, differ fundamentally from nanoconfined electrolytes sandwiched between chemically inert confining walls (e.g., graphene sheets)~\cite{Fong2025,Kong2017,Dockal2019}. 
 
A defining feature of  MnO$_2$ and similar materials is the presence of interlayer water. Confined water simultaneously screens electrostatic interactions, stabilizes and controls interlayer spacing, and reorganizes ionic coordination environments~\cite{Cheng2021,Luong2023,Johnson2006,MANG2022132293}. Consistent with this picture, studies on polycrystalline birnessite have identified intercalated water as the primary mediator of capacitive behavior, while spatially inhomogeneous ion distributions are found to promote local distortions of the MnO$_2$ sheet structure~\cite{Luong2023,nature_shelby2021}. Further, the interlayer water simultaneously stabilizes the structure and regulates capacitive behaviour through hydration-dependent lattice expansion and electrostatic screening~\cite{nature_shelby2021,Johnson2006,Shan2019}.

While hydration generally facilitates ionic mobility by reducing electrostatic binding in confinement~\cite{Shan2019,Merlet2012}, in recently synthesized MnO$_2$ single crystals, where ions were confined within layers of approximately $d_z \approx 7\mathrm{~\AA}$ inter-layer spacing,  a non-monotonic dependence of ionic conductions on water content emerged;  both strongly dehydrated and highly hydrated states exhibit suppressed transport across the micron-size MnO$_2$  flakes with a thickness of several atomic layers ~\cite{Parsi2025,serkan_hoca2021} (Figure~\ref{Fig:Intro}). Experiments have  further demonstrated nonlinear current--voltage characteristics, a hysteresis,  that can be controlled by water evaporation (i.e., vanishing hysteresis at elevated temperatures) in sodium-intercalated MnO$_2$ crystals~\cite{Parsi2025,serkan_hoca2021}. Given that the presence of confined water controls not only electrostatic screening but also the distance between confining surfaces in MnO$_2$ layers ~\cite{Chen2023NanoLetters,nature_shelby2021,Matsui2009,Li2020Nanotech}, it is highly likely that ion and water distributions affect the layer morphology under nonequilibrium  conditions via an unresolved microscopic mechanism.

Recent studies of nanoconfined electrolytes have emphasized that ionic transport can become inherently collective under strong confinement, leading to violations of the Nernst--Einstein relation and an emergent correlated motion~\cite{Fong2025,Robin2021,Angeles2023,Bocquet2025}. In layered metaloxides, an additional level of complexity arises because the confining host is not rigid and has a net charge. As a result, local variations in hydration may induce lattice distortions via heterogeneous electrostatic screening, which feeds back on ionic motion, potentially generating spatially heterogeneous transport pathways. Coupling between layer morphology, solvent structure, and transport has also been shown to strongly influence frictional and flow properties in confined hydrogen-bonded systems~\cite{Ma2015,Erbas2012}.

Here, we investigate field-driven ionic transport in sodium- and water-intercalated layered MnO$_2$, a self-intercalated metal oxide nanofluidic system, using nonequilibrium all-atom molecular dynamics simulations (NEMD). Our approach explicitly captures ion--water correlations together with the morphological response of atomically thin MnO$_2$ sheets under applied electric fields. We demonstrate that electric-field bias induces spontaneous nanoscale segregation of water within the interlayer space, producing coexisting hydrated and weakly hydrated ionic domains coupled to local lattice distortions. This feedback between hydration and morphology gives rise to heterogeneous transport pathways and leads to a non-monotonic dependence of ionic conductivity on water content.
By systematically separating the effects of hydration and layer flexibility, the non-monotonicity emerges from competing mechanisms: water-mediated expansion of interlayer spacing promotes transport, whereas excessive hydration stabilizes collective ionic structures that suppress ionic transport. Maximum conductivity occurs in an intermediate regime where partially hydrated boundary regions percolate across the layered crystal and dominate directional ion motion. Our results provide a microscopic mechanism linking confinement-induced electrostatic correlations, solvent organization, and structural dynamics to nonlinear ionic transport in layered metal-oxide vdW materials.

\section{Methodology}
\subsection{Simulation details}

In molecular dynamics (MD) simulations, all atoms in the system, including water molecules, ions, and the MnO$_2$ layer, are explicitly modeled using LAMMPS\cite{LAMMPS}. The MnO$_2$ sheets are prepared with replication of MnO$_2$ unit cell in all directions based on the experimentally reported lattice parameters ({Figure~\ref{Fig:Intro}a})~\cite{lopano2005}. A single unit cell is first replicated in the lateral (i.e., $\hat{x}$  and $\hat{y}$) directions to obtain a single Mn$_{900}$O$_{1800}$ + Na$_{225}^+$ nanosheet. Following,  these sheets are stacked in the $\hat{z}$ direction to obtain a total of 4 layers. Additionally, we tested the cases, in which 8 layer and did not observe any qualitative/quantitative difference in related metrics (see Supplemental Material~\cite{SM}, Fig.~S8). The layers are periodic in the lateral directions, and there is no structural irregularity between neighboring sheets. The layers at the top and bottom also confine ions to satisfy the periodicity in the $\hat{z}$ direction ({Figure~\ref{Fig:Intro}b}). The dimensions of the simulation box are $86.8\times75.2\times28.0$ \r{A}$^3$.

The pairwise interaction parameters between all atoms are defined \textit{via} the  CLAYFF forcefield~\cite{Cygan2021,Newton2020}. In this forcefield, the MnO$_2$ structure is stabilized by non-bonded interactions.   Monovalent sodium ions are chosen to model cations in accordance with the previous experiments~\cite{serkan_hoca2021,Parsi2025}. For Mn atoms constituting the sheets, an atomic charge of $q_{Mn}=+3.75e$, where $e$ is the elementary charge, is used to satisfy the mixed valence model of Mn$^{+4}$ and Mn$^{+3}$ valencies, upon modification of Lennard-Jones (LJ) pair coefficients~\cite{NEWTON2018208}. Surface oxygen atoms also carry partial charges  ({Supplementary Table S1}), and together with ions, they charge-neutralize the overall structure.  The total number of Na$^+$ ions $N_c$ is fixed in all simulations.
The short-range electrostatics between charged species are calculated via Coulomb potential with a cutoff distance of 10 \r{A} while the long-range electrostatics is calculated via the particle-particle particle-mesh method with an accuracy of 1.0e-4\cite{pppm}.  
The SPC/E water model is used in all simulations and is added randomly between sheets at prescribed concentrations.
These concentrations are referred to as $n$ H$_2$O$\cdot$Na, where $n$ is the number of water molecules per ion and varied between $n=0$ and $n=3$ throughout this work. 
Intrinsic water intramolecular bond lengths and bond angles are constrained to fixed values using the SHAKE algorithm in LAMMPS, eliminating high-frequency vibrational modes while preserving correct translational and rotational movements~\cite{shake}. The software package Packmol is used to design the layered surfaces filled with the prescribed amount of water molecules, which are initially randomly distributed~\cite{packmol2009}.

All simulations are run at a constant temperature $T$ and a constant number $N$ of atoms in the ensemble. In simulations where the surface atoms are thermalized, a constant-pressure ($P$) ensemble (i.e., $NPT$) is used to allow surface fluctuations. In $NPT$ calculations, the pressure is set to  $P=1$ atm in all directions by keeping the simulation box angles constant by using an anisotropic barostat~\cite{parrinello}. In simulations, where the inter-layer distance is fixed, the total volume of the simulation box $V$ is preserved to satisfy the $NVT$ conditions.  In $NVT$ simulations, the inter-sheet distance {$d_z$ is set to values ranging between $6.4 \leq d_z \leq 7.4$ \r{A}, and only ions and water molecules are allowed to move under the Nose-Hoover thermostat~\cite{tuckerman2006,shinoda}.

In the equilibration simulations preceding the application of the external electric field, ions and water molecules are energy minimized. Following this, regardless of the production ensemble used, an equilibration procedure is applied at $ T=300$ K with a damping constant of 100 fs and a time step of $\Delta t = 1$ fs for 20 ns. All data-production runs are performed for up to 100 ns with the same time step $\Delta t$ and temperature- and pressure-damping parameters under a constant electric field $E_x$ applied in the only  $\hat{x}$ direction (Figure \ref{Fig:Intro}b). The thermostat is applied only after excluding the center-of-mass movement contributions of the electric field. The maximum strength of the applied electric field is chosen so that the thermostat maintains the system temperature at the prescribed value throughout the simulation. 

\subsection{Calculation of ionic conductivity}
In order to characterize the ionic current under external electric fields, we calculate the average current from the current density $j$ as 
\begin{equation}
I_x =j A_{xy},
\end{equation}
where the $j= N q \langle v_x \rangle / A_{xy} d_z$. Here, $q = +1e$ is the charge of a monovalent sodium ion with $e$ being the elemantraty charge, $\langle v_x \rangle$ is the time and ensemble averaged velocity of all ions in the $\hat{x}$ direction,  $A_{xy}$ is the cross-sectional area of the interlayer channel, and $N_c$ is the number of ions in the simulation box.  Average interlayer spacing $d_z$ is calculated from the difference between the mean Mn positions of the two adjacent layers averaged over all layer pairs, and the error bars are the mean of the per-layer variances.
The velocity is calculated as the slope of the displacement data for the last 20 ns of each 100-ns-long simulation trajectory. The interlayer distance in $NPT$ simulations is obtained by averaging the vertical positions of all Mn atoms of each sheet and subtracting the mean values of adjacent layers. The standard errors are calculated and not shown if error bars are smaller than the symbol size. 

To calculate the steady-state average ionic velocity, ionic displacement averaged over all ions  is fitted with a function
\begin{equation}
\langle \Delta x \rangle = (1-\exp(-t/\tau))+ \langle v_x \rangle t.
\label{eq:Deltax}
\end{equation}
where $\tau$ defines a relaxation timescale, above which a linear increase in the displacement data is observed.
The exponential term in  Eq.~\ref{eq:Deltax} captures the rapid (i.e., $\sim$ 15 ns) response of the layer and molecular distributions to the electric field, while the second term describes the linear growth with a slope equivalent to the average drift velocity of ions at longer times (i.e., $t\gg \tau$). 

In our simulations, with 100-ns timescales accessible to MD, the electrical field strength is larger than that used in the experiments, which is, in fact, a common issue in non-equilibrium simulations~\cite{Gullbrekken2023,Netz2020Review,Erbas2012,Angeles2023,Rief2002}. Nevertheless, in simulation, the field strength must be sufficiently large for drift displacement to be statistically resolvable from thermal diffusion. The steady-state drift velocities from Supplementary Table S2 $(v_x \approx 0.006–0.039\;$\AA /ns) confirm that the minimum field is at the threshold of statistical detectability rather than dominating the dynamics. Further, the physical validity of the simulation at these field strengths is supported by the structural stability of the system: the MnO$_2$ sheets, sodium ions, and water remain in a physically consistent configuration throughout all production runs, with no indication of structural decomposition, unbounded ion drift, or thermostat failure,  consistent with the reversible and non-destructive character of the experimentally observed phase transitions in Ref.~\cite{serkan_hoca2021}.

\begin{figure*} [t]
\includegraphics[width=16cm,trim=0.5cm 0 0 0]{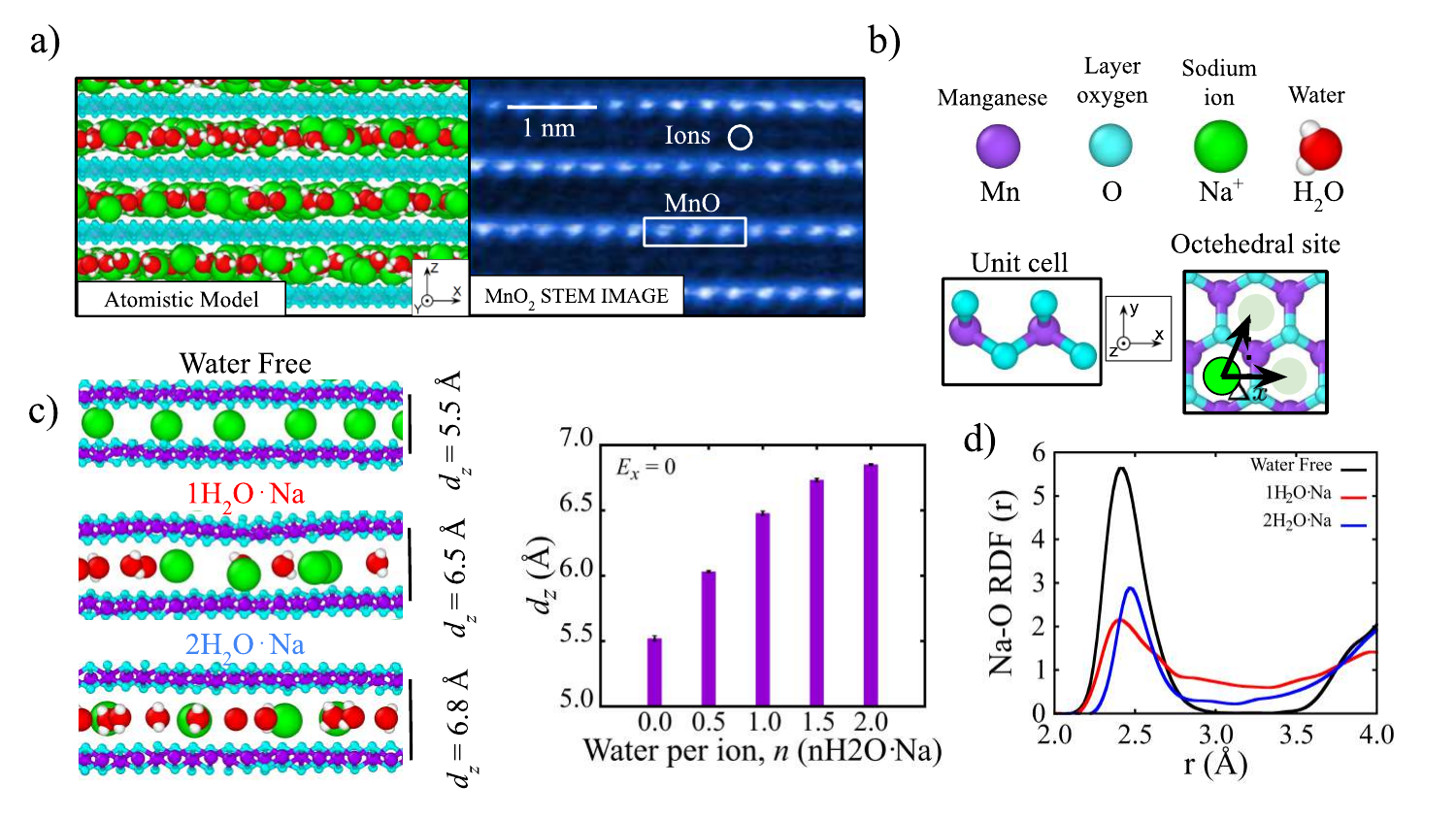}
\caption{Molecular-scale model of single-crystal MnO$_2$ layers: 
a) Representative MD snapshot and experimental image of a single-crystalline MnO$_2$ layer from the lateral ($\hat{y}$) direction. The experimental system is composed of 4 to 5 atomic layers~\cite{serkan_hoca2021}. The simulated model is constructed by stacking 4 MnO$_2$ sheets in the $\hat{z}$ direction. 
b) The atoms constituting the layers, including intercalated water and ions, are given in their vdW description. The unit lattice and the octahedral sites of the MnO$_2$ surfaces are shown with the same color codes.
c) Lateral view of the layers for various water levels and inter-layer thickness averaged over equilibrium (field-free) simulation trajectories. 
d) Sodium-layer oxygen (Na-O) radial distribution functions of electric-field-free cases for various water levels.}
\label{Fig:Intro}
\end{figure*}

\section{Results and Discussion}

\subsection{Electric field  bias alters water distribution across the vdW layers}

Inhomogeneous water distributions across MnO$_2$ vdW layers give rise to pronounced morphological distortions of the host lattice, manifested as spatial bending fluctuations and alterations in the thickness of the interlayer spacing~\cite{nature_shelby2021}. Consistent with these structural changes, experimental conductivity measurements reveal the coexistence of water-rich and water-poor domains under voltage biases, with significantly altered ionic mobility in less hydrated regions~\cite{Parsi2025,serkan_hoca2021}. To directly probe the microscopic origin of this phenomenon
we simulated layered MnO$_2$ sheets under external electric-field biases ranging between $E_x = 0$ and $E_x \simeq 10^{-2}$ V/\AA\  both under water-free and controlled-water-concentration conditions. These simulations explicitly account for the morphological response of the nanosheets to changes in the distributions of ions and water.

\begin{figure*}
\includegraphics[width=15cm]{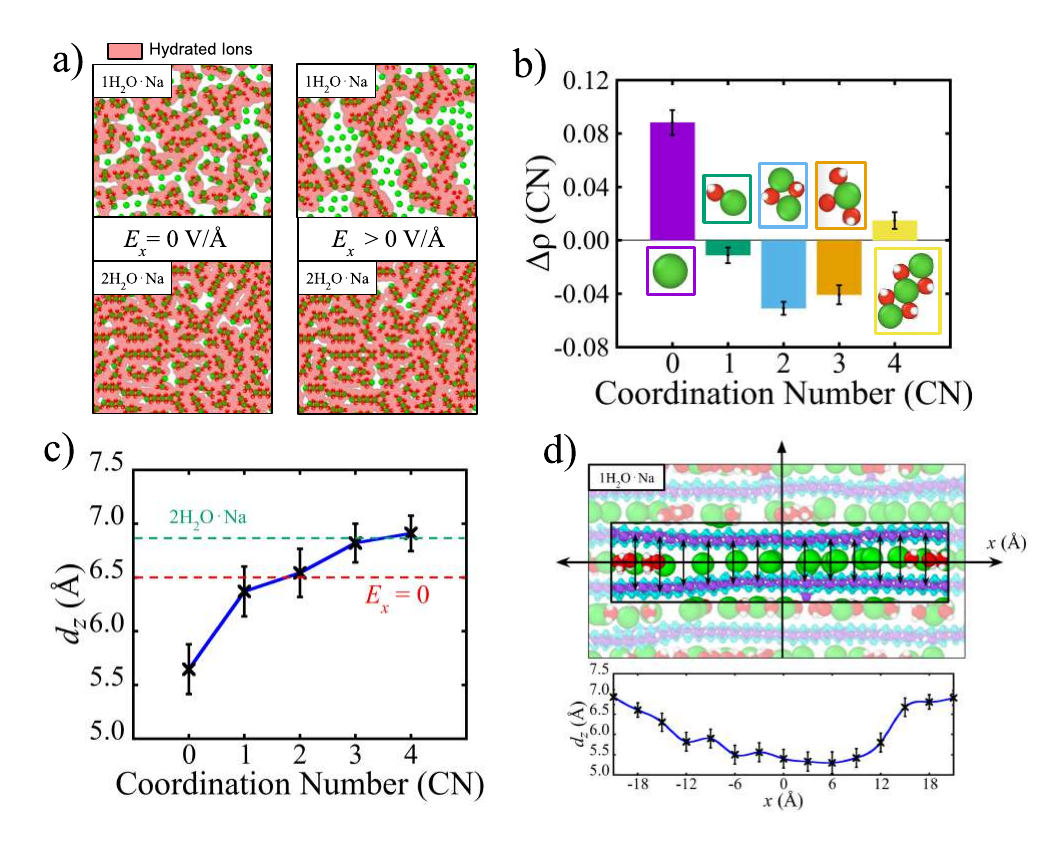}
\caption{Effect of electric field and water segregation and layer morphology:
a) Representative simulation snapshots for various water levels at the beginning and after 50 ns under an applied electric field.   
b) 
The change in the fraction of ions with a given coordination number (CN), defined as the number of water molecules in the first hydration shell, computed as the difference between equilibrium ($E_x = 0$) and the state at 50 ns under an applied field of $E_x = 0.02$ V/\AA, averaged over five independent simulations for the one-water-per-ion case (1H$_2$O$\cdot$Na). Snapshots show different coordination number configurations.
c) Average interlayer distance vs CN for 1H$_2$O$\cdot$Na.  
d) A snapshot demonstrating the undulations in the interlayer height and corresponding layer thickness after 50 ns at $E_x = 0.02$ V/\AA. 
}
\label{fig:npt_hop}
\end{figure*}

We first investigate the surface morphology of MnO$_2$ layers at varying levels of intercalated water, which can be experimentally controlled by vaporizing water at elevated temperatures~\cite{serkan_hoca2021}. To achieve this, we randomly distributed ions and water molecules across the layers and equilibrated the ionic positions for at least 20 ns. In the absence of water, MnO$_2$ sheets maintain their structural stability with an average, uniform interlayer spacing of $d_z \approx 5.5 \pm 0.2$~\AA, consistent with previous reports~\cite{Parsi2025,Cheng2021,Andre2024} ({Figure~\ref{Fig:Intro}} and  {Supplementary Figure~S1}). Notably, a fraction of Mn atoms occasionally diffuse toward the interlayer region; however, this doesn’t affect the overall stability within the time window of our simulations~\cite{cygan2012}.

Next, a prescribed number of water molecules per ion is introduced between the MnO$_2$ nanosheets ({Figure~\ref{Fig:Intro}}). Upon water intercalation, the average interlayer spacing increases markedly from $d_z \approx 5.5$~\AA\ to a water-concentration-dependent thickness ({Figure~\ref{Fig:Intro}c}). This expansion arises primarily from the screening of electrostatic attraction between the intercalated ions and the oppositely charged surface atoms by water molecules~\cite{Luong2023}({Figure~\ref{Fig:Intro}d}). The spatial distributions of water molecules and ions remain homogeneous over equilibrium simulations extending up to 20~ns in the absence of an external electric field, resulting in a uniform interlayer thickness across the sheets ({Figure~\ref{Fig:Intro}} and  {Supplementary Figure~S1}). Further increases in the water content beyond approximately two water molecules per ion lead to additional interlayer swelling~\cite{Fong2025} ({Supplementary Figure~S1}).

While water-intercalated vdW MnO$_2$ layers are structurally stable and exhibit a uniform interlayer spacing at zero electric field, the application of an in-plane electric field alters the coupled distributions of water molecules and sodium ions. Under non-equilibrium conditions, the field drives a lateral redistribution of both species, leading to the spontaneous formation of distinct nanoscale water--rich and water--poor domains across the sheets ({Figure~\ref{fig:npt_hop}a-b}). 
This domain formation is most pronounced when the water level is low, specifically when fewer than approximately two water molecules per ion are present, leaving a significant fraction of octahedral sites locally devoid of water, and progressively weakens at higher water concentrations ({Figure~\ref{fig:npt_hop}a}).

To quantify this effect, we characterize the local hydration environment using the 2D equivalent of Na--O coordination number (CN), where is the number of water molecules near the first hydration shell of an sodium ion ({Figure~\ref{fig:npt_hop}b-c}). This analysis reveals that while the population of ions with $CN=0$ and $CN=4$ (i.e., unhydrated and fully hydrated ions, respectively) increases, the number of ions with intermediate hydration levels decreases upon the application of an external field ({Figure~\ref{fig:npt_hop}b}).

The emergence of water-poor regions (with unhydrated ions) directly affects the host layer's morphology. Because water molecules screen the electrostatic attraction between intercalated sodium ions and the oppositely charged MnO$_2$ layers~\cite{Luong2023},  local solvent depletion strengthens ion--layer interactions. This enhanced attraction pulls the layers closer together, producing spatially varying interlayer distances and pronounced bending distortions of the sheets ({Figure~\ref{fig:npt_hop}c}). As a result, the interlayer spacing becomes explicitly dependent on the local hydration state ({Figure~\ref{fig:npt_hop}c-d}). For instance, near the ions with a hydration level of one water molecule per ion, the average interlayer distance decreases to $d_z = 6.2 \pm 0.3$~\AA\ due to reduced water-mediated screening in solvent-poor domains and further decreases near unhydrated ions ({Figure~\ref{fig:npt_hop}c-d and Supplementary Figure~S2}). 

Notably, once established, these hydration-driven structural inhomogeneities persist even after the external electric field is removed. The ionic and solvent distributions do not spontaneously re-homogenize within the accessible simulation time window, indicating either an effectively irreversible transition or a relaxation process that is slow compared to 2D ion and water diffusion. As the level of intercalated water is further reduced in the layers (i.e., less than 1 water per ion), the solvent-poor regions expand monotonically, and the interlayer spacing approaches the value observed in fully dehydrated layers, namely $d_z \approx 5.5$~\AA\ ({Supplementary Figure~S3}). 
To confirm that solvent segregation alone is sufficient to generate the observed structural distortions, we also performed control simulations in which water molecules were initially distributed non-uniformly in the absence of an electric field ({Supplementary Figure~S2}). These simulations reproduce the same hydration-dependent interlayer spacing and sheet distortions, demonstrating that the morphological response of the MnO$_2$ layers is primarily governed by local hydration rather than the electric field. Overall, our calculations demonstrate that voltage biases can induce water segregation, thereby affecting the interlayer thickness across the layers. As we will discuss next, this alteration can impact ionic movement across the MnO$_2$ vdW layers.

\begin{figure*}
\includegraphics[width=16cm]{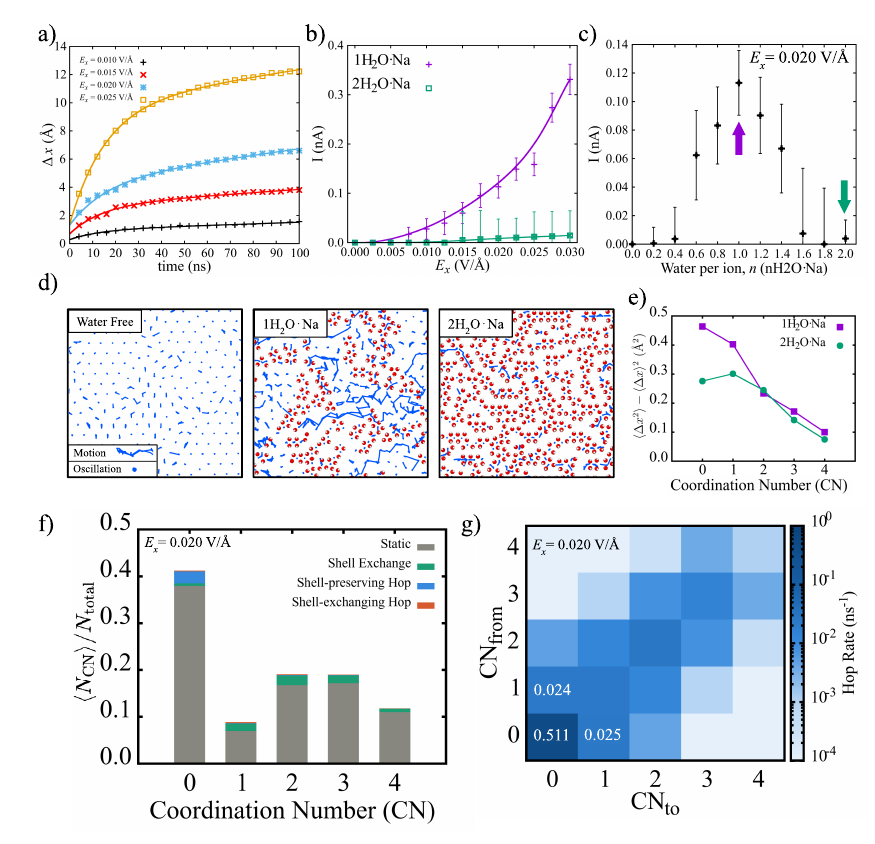}
\caption{Water segregation versus ionic conduction: 
a) The displacement trajectories averaged over all ions for the  1H$_2$O$\cdot$Na system for various electric field strengths. b) Electric field versus current for one and two water per ion cases, which correspond to the highest- and lowest-conducting systems in (c). 
c) The ionic current as a function of water levels at $E_x = 0.02\ \mathrm{V/\AA}$.  
d) Visualization of ionic trajectories for MnO$_2$ layers intercalated by various levels of water calculated at the steady state regime between 50 ns--100 ns interval of the simulation; blue lines denote the path of a single ion and blue dots refer to no motion. 
e) Coordination number vs. expected value of square of ionic displacement calculated over 1-ns-time windows for various water levels at $E_x = 0.02$~V/\AA. (f) Time-averaged fractional occupancy per coordination number, split into four categories by displacement and coordination change per frame. (g) Spatial hop rate matrix ns$^{-1}$ per ion for $\Delta x > 1$\AA\ events; diagonal entries are shell-preserving hops, off-diagonal are shell-exchanging hops, log color scale.}
\label{fig:disp_dist}
\end{figure*}

\subsection{Ionic motion is modulated by local hydration}
We reason that field-driven segregation of water into hydrated and unhydrated ionic domains across the vdW MnO$_2$ layers can have a direct, measurable impact on ionic transport dynamics (Figure~\ref{fig:disp_dist}). This expectation is consistent with voltage-sweep experiments, which show that above $\sim 400$~K -- where intercalated water desorbs from the MnO$_2$ single-crystal layers -- the pronounced current-voltage hysteresis observed at lower temperatures disappears~\cite{serkan_hoca2021}. This behavior suggests a transition to a transport regime in which ionic conduction is strongly suppressed in the absence of interlayer water.

Our simulations directly reproduce this water-dependent transition. In fully dehydrated MnO$_2$ layers, ions exhibit only initial displacement hops of less than $\sim 1$~\AA, even under electric field strengths as high as $E_x = 0.05$~V/\AA (i.e., comparable to the strengths to which water can break down). No sustained long-time ion motion is observed, resulting in effectively vanishing ionic conduction ({Supplementary Figure~S3}). This suppression reflects the combined effects of reduced interlayer spacing, enhanced ion-layer electrostatic attraction, and the absence of water-mediated screening.

By contrast, gradual intercalation of water into the layers expands the interlayer spacing, enabling field-assisted ionic transport ({Figures~\ref{fig:npt_hop} and \ref{fig:disp_dist}}). For hydrated layers, the average ionic displacement exhibits a characteristic two-stage temporal behavior: an initial exponential increase during the first $t \approx 15$~ns, followed by a linear increase at longer times. The crossover between these regimes coincides with the timescale over which water segregation develops within the layers  ({Supplementary Table 2}).  This indicates that the steady-state transport response emerges only after solvent redistribution is established in a field-dependent manner. 
Note that in experiments, the (macroscopic) electric-field sweeping rate is on the order of 1 mV/ms, and is much higher than used in simulations. Nevertheless, simulations probe the microscopic transport mechanism, in which a steady-state ionic drift-velocity regime is established after $\tau \approx 13-16$ ns (Figure~\ref{fig:disp_dist}a and Supplementary Table S2).

Steady-state ionic currents, extracted from the linear long-time displacement regime (see Methods), increase sub-linearly with applied electric field strength (Figure~\ref{fig:npt_hop}b). This deviation from Ohmic behavior indicates a field-dependent modification of ionic correlations and transport pathways, consistent with prior observations of non-Nernst-Einstein transport in confined electrolytes~\cite{Robin2021,Angeles2023}.

Because intercalated water simultaneously controls electrostatic screening and interlayer spacing, we next examine the dependence of ionic conductivity on the level of intercalated water (Figure~\ref{fig:disp_dist}c). Surprisingly, the conductivity exhibits a non-monotonic dependence on water content: both the lowest (e.g., 0.2 water molecules per ion) and the highest (e.g.,  2 water molecules per ion) water levels in our simulations lead to minimal ionic mobility, whereas the maximum conductivity occurs at intermediate water levels, ranging from approximately 0.5 to 1.5 water molecules per ion.

This non-monotonic behavior could be explained by two distinct mechanisms limiting the ionic motion: confinement versus excess intercalated water in the interlayers. At low water levels, the interlayers are partially collapsed, severely constraining ionic motion. At high water levels, the interlayer distance increases to $d_z \approx 6.8 \pm 0.1$~\AA\ and becomes spatially more uniform compared to the cases with less water. Nevertheless, ionic mobility is suppressed to levels comparable to the vanishing water cases (Figure~\ref{fig:disp_dist}b). Thus, in this strongly ion-hydrated regime, the reduced conductivity cannot be attributed to interlayer collapse.

At higher water levels (i.e., 1.5 water per ion or higher), water-poor regions vanish as most ions become fully solvated (Figures~\ref{fig:npt_hop}a and \ref{fig:disp_dist}d). We hypothesize that in this regime, water molecules occupy nearby octahedral sites that would otherwise serve as accessible hopping destinations for ions, thereby suppressing long-range ion motion (Figure~\ref{Fig:Intro}b and Supplementary Movie 1). Conversely, at very low water levels, ions are immobilized due to insufficient electrostatic screening and reduced interlayer spacing, which prevents hopping between neighboring octahedral sites. Notably, the water levels corresponding to maximum ionic conductivity coincide with the regime in which water-rich and water-poor 2D ionic domains largely coexist within the layers ({Figures~\ref{fig:npt_hop} and \ref{fig:disp_dist}}), suggesting a fine competition of the aforementioned effects.

Trajectory analysis further confirms that in the absence of water and with high water levels, ions cannot move along the interlayers (Figure~\ref{fig:disp_dist}d). At intermediate water levels, where there is co-existence of water--poor and --rich regions, a visual inspection reveals that ions with little or no local hydration move over significantly longer distances than fully hydrated ions (Figure~\ref{fig:disp_dist}d and Supplementary movie 2). Surprisingly, only ions residing near the boundary of water-poor zones contribute to directional ionic motion. At lower water levels (e.g., 0.2 water molecules per ion), extended water-poor ionic clusters span large regions of the layers, causing the interlayer spacing to approach the dehydrated limit locally ({Supplementary Figure S3})

To further quantitatively assess the role of water in ion transport, we compute the displacement histograms of 2D ionic jump lengths along the field direction for each hydration level ({Supplementary Figure S5}). The distribution of ionic displacements $\Delta x$ over time intervals of $\Delta t =$ 1 ns, and 10 ns shows a pronounced asymmetry towards ionic hops in the field direction, particularly for  CN$=0$ under applied fields  ({Supplementary Figure S5}).
Calculating expected values and variances for ion jumps from these distributions reveals that jump lengths are minimal at both low and high water contents, consistent with negligible conductivity in these regimes (Figure~\ref{fig:disp_dist}d-e and Supplementary Figure S5).  However, at intermediate water levels (i.e., one water per ion), ions exhibit longer hopping distances on average and longer individual ionic trajectories. In fact,  ions with CN~$=0$ contribute disproportionately to ionic hops, while ions with CN~$=4$ exhibit the smallest hops ({Figure \ref{fig:disp_dist}e}).

Taken together, these results demonstrate that ionic transport in intercalated MnO$_2$ layers requires a delicate balance: the interlayer spacing must exceed a critical threshold to permit hopping, while excessive hydration suppresses transport by blocking accessible hopping sites. Maximum ionic conductivity, therefore, emerges only in an intermediate hydration regime, where partial solvation enables electrostatic screening without eliminating available transport pathways.

\subsection{Weak hydration promotes faster ion kinetics in locally swollen regions}
In simulations, a visual analysis reveals that the number of mobile ions comprises only a small fraction of the total ions (Supplementary Movie 2). Hence, we next ask if all $\text{CN}=0$ ions contribute to the ionic current and performed a CN-resolved ion activity analysis (Figure~\ref{fig:disp_dist}f, g). This analysis reveals that transport in the 1H2O·Na system is dominated by a small fraction of ions. Each frame-pair is classified (Figure~\ref{fig:disp_dist}f) into one of four mutually exclusive categories based on whether the ion displaced more than 1 Å ($|\Delta x| > 1$ Å) and whether its coordination number changed ($\Delta\text{CN} \neq 0$) within a 0.05 ns window. The 0.05 ns window is chosen as approximately two orders of magnitude shorter than the mean inter-hop time ($\sim$10 ns per ion, derived from the total outgoing hop rates of order $10^{-2}$ ns$^{-1}$), ensuring that each frame-pair captures at most one transition event per ion. Ions satisfying neither criterion are classified as static  ($|\Delta x| \leq 1$ Å, $\Delta\text{CN} = 0$), meaning they neither displace nor exchange coordinating water molecules. Ions satisfying only the coordination criterion undergo local shell rearrangement ($|\Delta x| \leq 1$ Å, $\Delta\text{CN} \neq 0$), where the coordination shell reorganizes without associated spatial displacement, representing thermal fluctuations of water molecules across the coordination cutoff radius. Ions satisfying only the displacement criterion make shell-preserving hops ($|\Delta x| > 1$ Å, $\Delta\text{CN} = 0$), displacing between octahedral sites while retaining the same number of coordinating water molecules. Ions satisfying both criteria make shell-exchanging hops ($|\Delta x| > 1$ Å, $\Delta\text{CN} \neq 0$), displacing between octahedral sites while simultaneously gaining or losing a coordinating water molecule. At all coordination states, the majority of ions are static, with $\text{CN} = 0$ carrying the largest total occupancy ($\langle N_{\text{CN}=0} \rangle / N_\text{total} \approx 0.41$). Despite this, $\text{CN} = 0$ exhibits the highest shell-preserving hop rate (0.51 ns$^{-1}$ per ion), driven by its large population rather than elevated per-ion mobility. The mobility of these dehydrated ions is attributed to the local structural expansion induced by neighboring hydrated ions, whose water molecules increase the interlayer spacing and effectively lower the steric barrier for $\text{CN}=0$ ion displacement between sites. Among hydrated ions ($\text{CN} \geq 1$), the hop rate matrix shows that spatial displacement is predominantly accompanied by nearest-neighbor CN transitions ($|\Delta\text{CN}| = 1$), with rates of 0.012--0.025 ns$^{-1}$. Long-range jumps in coordination space ($|\Delta\text{CN}| \geq 2$) are at least one order of magnitude smaller, indicating that ions gain or lose coordinating water molecules one at a time rather than through direct multi-step transitions. Forward and backward rates between adjacent CN states are nearly symmetric, indicating that the field does not significantly bias the hydration-shell exchange process at this field strength.

\begin{figure*}
\centering
\includegraphics[width=15cm]{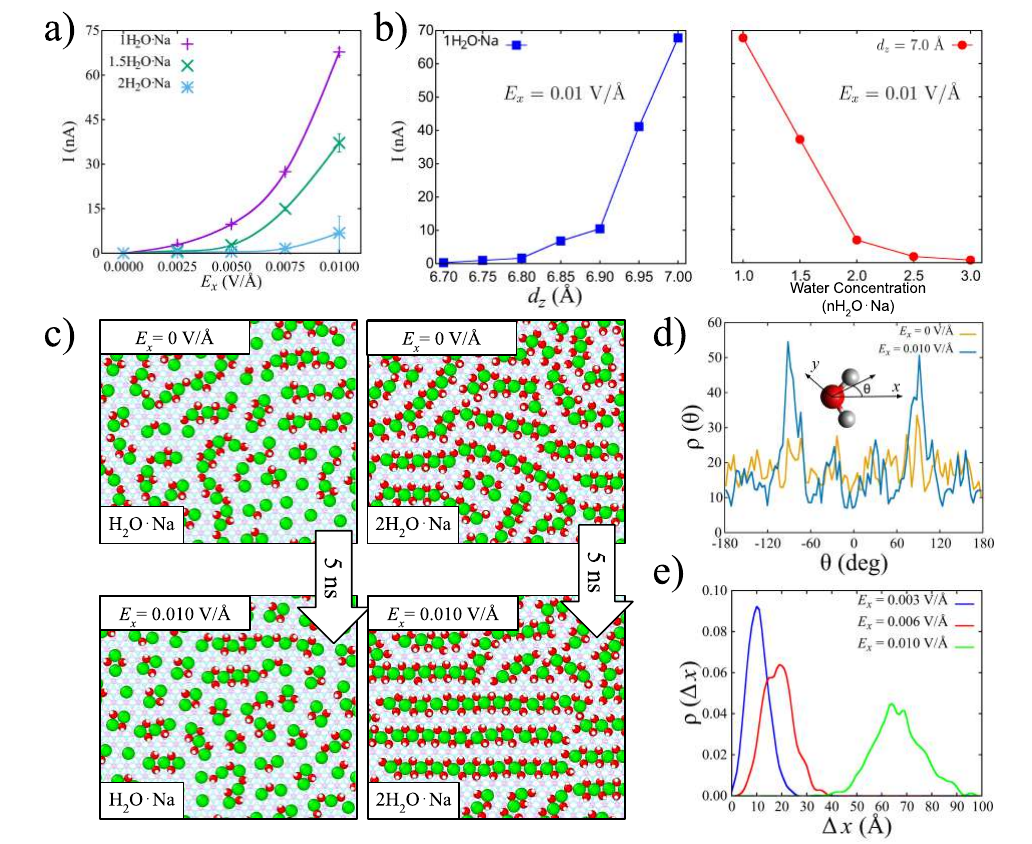}
\caption{Ionic transport in rigid, fluctuation-free vdW layers. 
a) Applied electric field versus current for various water levels. 
b) Ionic current for (left) various interlayer distances at a fixed water level and  (right) for various water levels at a fixed interlayer distance. 
c) Representative simulation snapshots demonstrating the ion and water distribution in a single layer ($d_z = 7 \ \mathrm{\AA}$) of MnO$_2$ at equilibrium and under an applied field of $E_x = 0.01$ V/\AA. 
d) The distributions of the angle between the $x$ axis and the water dipole on the $xy$ plane.
e) Displacement histograms of ions at $d_z = 7 \ \mathrm{\AA}$ interlayer spacing for 10 ns-long  simulations at various field strengths for 1H$_2$O$\cdot$Na case.} 
\label{fig:nvt_hop}
\end{figure*}

\subsection{Non-monotonic ionic conduction is due to an interplay of hydration and layer swelling}

Our observation of a non-monotonic dependence of ionic current on intercalated water levels indicates the presence of two competing effects  (Figure~\ref{fig:disp_dist}c). At high levels of interlayer water, water molecules hinder ion transport by occupying migration pathways and stabilizing ions in their octahedral sites, thereby suppressing hopping events, as previously reported in confined and layered systems~\cite{Cheng2021,Shan2019,Yoshisako2025}. 
In contrast, at lower hydration levels, water plays a beneficial role by screening electrostatic interactions between ions and layer atoms, thereby non-uniformly expanding the interlayer spacing and facilitating the motion of partially or fully water-free ions.

To disentangle these two effects, we perform separate simulations, in which the MnO$_2$ sheets are treated as rigid, static layers with a fixed interlayer spacing $d_z$. 
This frozen-layer approximation, commonly used in studies of confined electrolytes~\cite{Angeles2023,Robin2021,Fong2025}, eliminates morphological fluctuations of the host lattice and allows us to isolate the effects of interlayer spacing and hydration on ionic motion independently~\cite{Shan2021,Ma2015,Erbas2012}. 
Within this framework, we systematically vary the interlayer distance while maintaining a fixed spacing and, in separate simulations, independently vary the water content while keeping the interlayer distance fixed.

We first set out to calculate the ionic current for a fixed interlayer distance of $d_z = 7.0$~\AA  for various water levels (Figure~\ref{fig:nvt_hop}a). Compared to the ionic conduction in flexible nanosheets (Figure~\ref{fig:disp_dist}b), ions can move much faster in interlayers whose lattice fluctuation is absent. Notably, unlike the deformable layers, the rigid layers do not exhibit an exponential increase in the ionic displacement, indicating the absence of solvent segregation  ({Supplementary Figure S6}). These simulations suggest that bending and distortion of the layers can further confine the ions and limit their motion. 

For the water level, at which we observe the maximum ionic conduction in unconstrained layers, namely, one water molecule per ion, we vary the interlayer spacing, the ionic conduction decreases at separations $d_z \leq 7.0$~\AA  and becomes unmeasurable below  $d_z = 6.0$~\AA in simulations~(Figure~\ref{fig:nvt_hop}b-c).  
This confinement-induced suppression of mobility at small separations is consistent with trends reported for monovalent electrolytes confined between graphene sheets~\cite{Fong2025}. 
These results also explain why, in unconstrained MnO$_2$ layers, gradual hydration enhances ionic conduction by expanding the interlayer distance and reducing ion hopping at moderate water levels  (i.e.,~one water per ion in Figure~\ref{fig:disp_dist}).

Next, to isolate the role of water levels from the inter-layer thickness, we introduce controlled amounts of water---up to three water molecules per ion---between sodium-intercalated MnO$_2$ layers while maintaining a constant interlayer spacing of $d_z = 7.0$~\AA~(Figure~\ref{fig:nvt_hop}b-c). 
In equilibrium simulations without an external electric field, ions remain strongly bound to their initial octahedral sites, and thermal fluctuations alone are insufficient to cause ion diffusion within the duration of our simulations    (Supplementary Movie 3). 

Once the electric field is switched on in simulations,  the ionic conduction is zero in completely water-free layers up to an electric-field-dependent threshold.
Contrarily, in the presence of intercalated water, less hydrated systems (e.g., one water molecule per ion) consistently exhibit higher ionic conductivity than more hydrated systems (two water molecules per ion) over a broad range of electric field strengths  (Figure~\ref{fig:nvt_hop}b). 
We attribute the reduction of ionic mobility as the intercalated water levels is increased to the formation of stable, 2D hydration shells around ions: at two water molecules per ion, each ion is coordinated on average by four octahedrally bound water molecules, which strongly localize the ions and suppress their hopping  (Figure~\ref{fig:nvt_hop}c and Supplementary Figure S6). Representative simulation snapshots reveal that, under fixed interlayer spacing, ions organize into filamentous clusters aligned along the field direction  (Figure~\ref{fig:nvt_hop}c). 
At high water levels, these ion ``trains'' are highly ordered, where water dipoles are aligned perpendicular to the field direction to stabilize these trains ({Figure~\ref{fig:nvt_hop}c,d}). 
These well-hydrated ion trains exhibit minimal motion despite drifting collectively in the field direction, resulting in negligible net displacement and low conductivity (see Supplementary Movie 4). 
Correspondingly, the average ion--ion distance decreases under applied fields in the highly hydrated case, indicating enhanced electrostatic screening that promotes clustering  ({Supplementary Figure S6}).
Notably, repeating simulations with extended lateral dimensions yield similar filamentous cluster formation patterns. 
At sufficiently strong electric fields (i.e.,$E_x \gtrsim 0.01$~V/\AA), the clusters span the entire simulation box along the field direction (Supplementary Figure~S6), indicating that their growth is not an artifact of system size.

In contrast, at lower water levels, confined ions have missing water at their hydration shells (e.g., less 4 water per ion) and some octahedral sites remain unoccupied by water, providing accessible hopping destinations for ions. 
Under these conditions, ionic clusters are shorter, structural ordering is reduced, and ionic mobility is enhanced  (Figure~\ref{fig:nvt_hop}). Importantly, the field-induced alignment of ions with water molecules persists even at larger interlayer separations in rigid, fluctuation-free layers  ({Supplementary Figure S7}),

Overall, simulations performed under a constant interlayer spacing with varying water levels confirm two opposing mechanisms governing ionic transport in MnO$_2$ layers. 
Interlayer expansion promotes conduction by reducing confinement and electrostatic binding, while excessive hydration suppresses conduction by stabilizing ions in water-coordinated clusters that block hopping pathways. The formation of finite-size water-rich and water-poor domains favors the optimum thickness and hydration levels for ionic motion.  
The competition between these effects could explain the experimentally observed non-monotonic dependence of ionic conductivity on hydration level.

\section{Conclusion}

Water plays a dual and indispensable role in stabilizing ion-intercalated MnO$_2$ vdW-layered structures and enabling ionic transport within layered structures. 
In the absence of interlayer water, MnO$_2$ sheets collapse, are distorted, and lose ionic conductivity, consistent with prior experimental and theoretical studies~\cite{Johnson2006,Shan2019,Parsi2025}. 
Negatively charged MnO$_2$ surfaces strongly interact with confined water molecules, orienting their dipoles and disrupting the bulk hydrogen-bond network~\cite{Remsing2015}, while simultaneously regulating the interlayer spacing that governs ionic mobility.

In this work, inspired by previous experimental findings, we combined non-equilibrium all-atom MD simulations to elucidate the mechanisms governing field--driven ion transport in monovalent-ion-intercalated MnO$_2$ layers. 
Our results demonstrate that ionic conduction arises from a subtle interplay between ion hydration, electrostatic screening, and the water-mediated morphology of the confining vdW layers. 
Water molecules stabilize Na$^+$ ions by forming hydration shells, which can suppress ionic hopping, in agreement with previous studies of confined electrolytes~\cite{YiHeng2021}.

When MnO$_2$ sheets are allowed to deform in response to local water levels, water becomes essential for stabilizing the layered morphology and preventing the formation of current-dead regions. 
Our simulations reveal that electric fields induce spatial segregation of water into water-rich and water-poor domains, leading to nanoscopic regions occupied by hydrated and unhydrated ions, respectively. This segregation also results in a hydration-dependent interlayer spacing and heterogeneous ionic transport pathways  (Figures~\ref{fig:npt_hop} and \ref{fig:disp_dist}).  Specifically, ions located at the boundaries of these domains contribute most to conduction, while strongly hydrated ions are minimally mobilized by the external field.
If boundary regions percolate across a finite-size layered crystal, ionic conduction is sustained; if not, conductivity can vanish entirely. 
Consequently, we observe a monotonic decrease in ionic conductivity with increasing water levels, consistent with experimental measurements on micron-scale MnO$_2$ crystallites  (Figures~\ref{fig:disp_dist} and \ref{fig:nvt_hop}).

Notably, although field-induced segregation occurs only for flexible layers, rigid layers lead to markedly higher ionic currents. This contrast highlights the morphological response of the host lattice to the external voltage biases. 
Surface oscillations have previously been shown to strongly influence molecular friction and water transport~\cite{Erbas2012,Bocquet_Netz_2015}. Our simulations also demonstrate that the absence of atomic lattice oscillations (i.e., no phonons) triggers a collective ion-hopping mechanism, thereby enhancing conductivity quantitatively (Figures~\ref{fig:disp_dist} and \ref{fig:nvt_hop}).  Thus, surface dynamics may emerge as a key control parameter for optimizing ionic transport in ion-intercalated layered materials.

Our computational findings can also shed light on the experimentally observed disappearance of $I-V$ hysteresis at high temperatures~\cite{serkan_hoca2021}. At elevated temperatures, intercalated water evaporates, leaving the layers water-free. With vanishing intercalated water, electrostatic screening decreases, and the interlayer distance between  MnO$_2$ sheets decreases, in accord with our simulations with a water-free layer.  In the water-free layers, applied voltage does not change the morphology of the layers or cannot lead to water-rich or water-poor surface regions. Consequently, the reversal of voltage direction also causes no change in the layers. Thus, in the absence of intercalated water, $I-V$ hysteresis disappears~\cite{serkan_hoca2021}.
Conversely, as the temperature is lowered, water diffuses back into the interlayers, and the hysteresis reappears reversibly. We reason that intercalated water disrupts the ionic/water distribution within the layers and consequently distorts layer morphology during a forward voltage sweep. Once the voltage direction is reversed, ions need to rearrange the clusters in water — into poor and rich regions — and possibly reset the layer morphology, which can lead to a lower current response than in the forward sweep.

The forcefield used in this study, CLAYFF, was directly validated for phyllomanganate systems by Newton \& Kwon~\cite{Newton2020,NEWTON2018208}, reproducing interlayer spacings and Na$^+$ coordination environments consistent with our Na–O radial distribution function (Figure~\ref{Fig:Intro}d).  Nevertheless, under an applied electric field, polarization effects, which are not included in CLAYFF, may be important. However, we reason that polarization effects may not qualitatively alter the conclusions of this study for the following reasons. First, the transport mechanism identified here is geometric rather than electrostatic. The coordination-number-resolved displacement analysis (Figure S5c) shows that transport is carried exclusively by CN = 0 ions adjacent to vacant octahedral sites, while fully coordinated ions remain immobile. Concurrently, the polarizability of Na$^+$ is known to decrease significantly upon confinement from its free-ion value, which is already small~\cite{coker2002}. For the Mn species, any polarization correction would be spatially uniform across all equivalent octahedral sites since Mn atoms sit in a regular crystallographic environment. A spatially uniform correction modulates absolute site energies but should not alter which sites are occupied or vacant, possibly leaving the CN-resolved transport mechanism unchanged. In fact, the dominant field-response in the interlayer is reorientation of the water permanent dipole ($\mu$SPC-E = 2.35 D)~\cite{berendsen2002}, whose energy $U = \mu E_x \approx 1\ kBT$ at $E_x = 0.05$ V/\AA\, suggesting that at the strongest field used here, even water dipoles do not align along the field direction, demonstrating that confinement forces dominate over field-induced orientational. Therefore, we expect negligible polarization responses for other constituting species.

Finally, stacked MnO$_2$ layers exhibit strong dielectric anisotropy, characterized by high in-plane and low out-of-plane permittivity~\cite{Frisenda2020,Arias2019,Moreira2001}.  This strong screening along the direction perpendicular to the surface effectively decouples adjacent layers, amplifying the influence of local morphology and hydration on ionic motion. 
Taken together, our results demonstrate that ionic conduction in MnO$_2$ vdW layers is governed not solely by ion hydration or confinement, but by their collective coupling to dynamic surface morphology and electrostatic correlations. 
These insights can provide a rationale to tune ionic transport in voltage-controlled ionic devices composed of intercalated vdW structures.

\begin{acknowledgments}
The numerical calculations reported in this paper were partially performed at TUBITAK ULAKBIM, High Performance and Grid Computing Center (TRUBA resources). 
AUO acknowledges the  TUBITAK  124N935 and TUBITAK National PhD Scholarship Program 2211. 
\end{acknowledgments}

\bibliography{apssamp}

\clearpage
\onecolumngrid
\newcount\SMpg \SMpg=1
\loop\ifnum\SMpg<11
  \noindent\includegraphics[page=\the\SMpg,width=\textwidth,%
      height=\textheight,keepaspectratio]{Supplementary_Information_APS.pdf}%
  \clearpage
  \advance\SMpg by 1
\repeat

\end{document}